\documentclass{article}
\usepackage{spconf,amsmath,graphicx, color}
\usepackage{soul}

\title{A novel 1D state space for efficient music rhythmic analysis}
\name{Mojtaba Heydari$^{\ddagger \dagger}$ \qquad Matthew McCallum $^{\star}$ \qquad Andreas Ehmann$^{\star }$ \qquad Zhiyao Duan$^{\dagger}$ \thanks{$\ddagger$ Main work is accomplished as a research intern at Pandora Media Inc.  \ This work is partially funded by National Science Foundation grant No. 1846184.}}
  
\address{$^{\dagger}$ University of Rochester, Rochester, NY, USA \hspace{0.5in}
      $^{\star}$ Pandora Media, Inc., Oakland, CA, USA\\ \tt \normalsize  mheydari@ur.rochester.edu}

\begin{document}
%
\maketitle
\begin{abstract}
Inferring music time structures has a broad range of applications in music production, processing and analysis. Scholars have proposed various methods to
analyze different aspects of time structures, such as beat, downbeat, tempo and meter.
Many state-of-the-art (SOFA) methods, however, are computationally expensive. This makes them inapplicable in real-world industrial settings where the scale of the music collections can be millions. This paper proposes a new state space and a semi-Markov model for music time structure analysis. The proposed approach turns the commonly used 2D state spaces into a 1D model through a jump-back reward strategy. It reduces the state spaces size drastically. We then utilize the proposed method for causal, joint beat, downbeat, tempo, and meter tracking, and compare it against several previous methods. The proposed method delivers 
similar performance with the SOFA joint causal models with a much smaller state space and a more than 30 times speedup.
\end{abstract}
\section{Introduction}
\label{sec:intro}

Time is a fundamental concept in music. Automatic analysis of music time structures enables many applications in music generation, manipulation, and recommendation. As an example, such analysis is essential for tasks like music-score alignment ~\cite{cont,duan:1} and music transcription ~\cite{shibata}. Several approaches have been proposed to extract music rhythmic parameters such as tempo, meter, beat, and downbeat. Although SOFA models for some of the mentioned tasks are promising, computational efficiency considerations make many of them not applicable in massive industry-scale settings. Furthermore, recent advancements of augmented and virtual reality and their interactive applications demands real-time processing, in which computational efficiency is a critical factor.     

Many models attempt to estimate music rhythmic parameters separately e.g.,~\cite{Durand:1,Heydari,Oliveria} . However, due to the underlying inter-dependencies among these parameters, joint approaches have demonstrated promising results as well~\cite{Bock:1,Heydari:2,Peeters:1}. Joint models often employ state spaces to model these inter-dependencies, and use some probabilistic models such as Bayesian models to infer the parameters of interest. As rhythmic parameters are intrinsically continuous variables, the state spaces are ideally continuous or discretized with a fine granularity. Hence, the inference can be computationally expensive when the state spaces are high dimensional.


To make the inference stages tractable, some works e.g.,~\cite{duan:1,Srinivasamurthy:1,Hainsworth:1,Cemgil:1,Holzapfel:2} used numeric approaches such as particle filtering (PF) to approximate the probability of the states in continuous state spaces. Another set of approaches e.g.,~\cite{whiteley:1,Krebs:1,Peeters:1,krebs:2,Holzapfel:1} discretize the state space to deal with a limited number of hidden variables during the inference. Discretizing the state spaces makes it possible to use some efficient inference algorithms
such as the forward algorithm to compute 
posterior probability of the hidden states. 

In our previous work~\cite{Heydari:2}, we demonstrated that the combination of the ideas mentioned above can deliver the SOFA performance for causal and joint time structure analysis. It utilized a cascade of discrete state spaces and an enhanced PF inference strategy using a proposed information gate technique to make the inference faster. 

In contrast to the above-mentioned Bayesian methods that assume the Markovian property, in this paper, we introduce a compact 1D state space and a semi-Markov jump-back reward technique where the transition model is time-variant. Such variance across time is likely to make the inference process more complex, in return, it makes the model much more efficient. Although semi-Markov processes are used before for some specific tasks, e.g., score-alignment ~\cite{cont}, here we propose a generic model that can be utilized as an efficient alternative for the well-known bar pointer models to improve the efficiency on several music rhythmic analysis tasks. 


\section{Approach}
\label{sec:approach}

In this section, we describe the jump-back reward strategy. To elaborate it, we first describe the bar pointer model~\cite{whiteley:1} and its more efficient version~\cite{Krebs:1}. Then, we present the new model and demonstrate how it covers the same parameters i.e., tempo range and time resolution, with much fewer states.

\subsection{Bar pointer models}
\label{ssec:bar pointer model}
The \textit{bar pointer model} for joint meter, tempo, and rhythmic pattern inference 
was proposed by Whiteley et al.~\cite{whiteley:1} to model the dynamics of a bar pointer that moves through a 2D state space throughout a piece of music. The two dimensions of the state space represent relative positions within a bar $\Phi_{k}\in\{1,2,...,M\}$ with $M$ being the last position, and the tempo (in the unit of frames per bar) $\dot\Phi_{k}\in\{ \dot\Phi_{min},\dot\Phi_{min+1},...,\dot\Phi_{max}\}$, in the $k$-th audio frame. This state space is discretized into even grids along both dimensions, assigning the equal number of positions per bar for different tempi, leading to different time resolutions for them.

Under the same framework, assuming a known and fixed meter, Krebs et al.~\cite{Krebs:1} proposed a different discretization strategy for the state space, where the number of positions within a bar $M(T)$ depends on the tempo $T$ (in BPM) as:
\begin{equation}
M(T)= Round (\frac{ B\times 60 }{T\times\Delta}),
\end{equation}
where $B$ is the number of beats per bar that is fixed and known, and $\Delta$ is the frame length in seconds. By assigning fewer positions for higher tempi, this model ensures 
%
the same time resolution for different tempi.
This makes the inference more efficient hence is called the \textit{efficient bar pointer model}. 

Similarly, for beat tracking alone, a \textit{beat pointer model} and its efficient counterpart can be derived, by replacing the horizontal axis with the relative positions within a beat.


When the meter is unknown, however, one limitation of the efficient bar pointer model is that 
the inference needs to be performed on multiple state spaces \textit{independently}, making the total state space much larger.
For instance, given the tempo axis ranging from 55 BPM to 215 BPM and the audio frame hop size of 20 ms, if a bar may contain 2 to 6 beats, then a total of 28,980 states will be needed counting all of the 5 independent state spaces. This problem is addressed by some works such as the BeatNet~\cite{Heydari:2}, using a cascade model of two separate state spaces for music beat-tempo tracking and downbeat-meter tracking, respectively. 
This allows to consider different bar lengths in a single state space, and shrinks the total state space size from 28,980 states down to 1,469 states for the example above. Fig. 1a and 1c demonstrate the state space of the \textit{efficient beat pointer model}. and the downbeat state space to cover a bar length from 2 to 6 beats for the cascade model. The disadvantage of the cascade model is that the inference of downbeat and meter depends on the inference of beat and tempo. In other words, errors in beat and tempo tracking will be propagated to downbeat and meter tracking.
 
\begin{figure*}[htbp]
 \centerline{
 \includegraphics[width=1.02\textwidth]{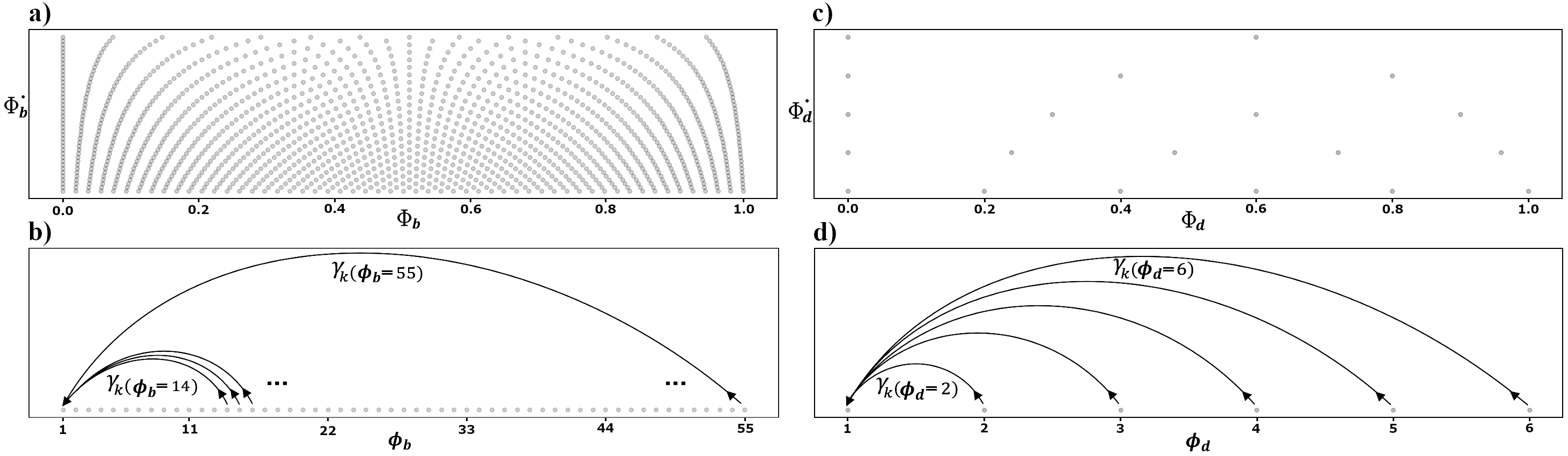}}
 \caption{Comparison of the state spaces for music time analysis for a tempo range of [55, 215] BPM, a frame hop size of 20 ms, and a bar length of [2, 6] beats. a) The efficient beat pointer state space ~\cite{Krebs:1} for beat-tempo tracking. b) The proposed 1D state space with jump-backs for beat-tempo tracking. c) The cascade state space for downbeat-meter tracking~\cite{Heydari:2}. d) The proposed 1D state space with jump-backs for downbeat-meter tracking.}
 \label{fig1}
\end{figure*}




\subsection{Proposed 1-D state space}
\label{ssec:1D state space}
The main idea of the proposed approach is to modify the 2D bar pointer or beat pointer state space into a compact 1D space that covers the same tempo or meter range and time resolution but with much fewer states. This is achieved by re-defining the position dimension and replacing the tempo dimension with a jump-back strategy. 

Specifically, taking the beat space as an example, instead of defining the position dimension as relative positions within a beat, the proposed 1D state space defines it as audio frame indices within a beat interval. The first index corresponds to the beat position, while the last corresponds to the last frame within the beat for the smallest possible tempo. As an audio frame arrives, a pointer hypothesis moves along this dimension one step to the right. It jumps back to the left end when it believes that a new beat arrives. As the tempo is unknown, there is uncertainty on when jump-back should happen. We hence use a probability distribution $\gamma(\phi_{k})$ to describe this uncertainty, where $\phi_{k}$ is the position (i.e., the frame index within a beat) of the $k$-th audio frame. This is illustrated in Fig.~1b. The key of this jump-back operation is that it circulates back some probability mass of the pointer's position distribution to the left end (i.e, the beat position), which is then gradually transported to the right as new frames arrive. When the music demonstrates regular beats, the jump-back probability will show a strong peak at the frame index corresponding to the beat interval, and the posterior distribution of the pointer position will be reinforced to show a significant peak that moves forward but circles back after each beat interval.
 
Fig.~1a illustrates the efficient beat pointer state space ~\cite{Krebs:1} that includes multiple rows 
corresponding to different tempi from $M(T_{min})=55$ (frames per beat interval) in the bottom row to $M(T_{max})=14$ (frames per beat interval) in the top one. The proposed state space, in contrast, is 1D as demonstrated in Fig. 1b. 
The number of states in Fig. 1b is equal to the number of frames within one beat interval for the smallest possible tempo. By taking the same example used in Fig. 1a, this number is 55. Therefore, the number of states is reduced from 1,449 in Fig. 1a to 55 in Fig. 1b. 

This 1D state space can also be constructed for the downbeat (bar) state space following the same logic. Fig. 1c shows the bar state space in the cascade approach ~\cite{Heydari:2}, where the horizontal axis is relative positions within a bar at the granularity of a beat, and the vertical axis is the bar length. Fig. 1d reduces this space into 1D with only the position dimension; the bar length range and the position granularity are kept the same as those in Fig. 1c. It is clear that the number of states reduces from 20 in Fig. 1c to 6 in Fig. 1d. 

In addition to assisting with computational cost reduction, fewer states may help increase the accuracy, given that the system should infer the correct states out of fewer hypotheses. Note that the new state space includes a vector of jump-back weights corresponding to the beat/bar positions, and the position that achieves the maximum one represents the local tempo/meter.
The weights are updated based on the reward-punishment mechanism discussed in the next session.

\subsection{Inference}
\label{ssec:Inference}
This section describes an approach to incorporating the proposed state space into the HMM process for online music rhythmic analysis tasks in which the inference cost and speed are crucial factors. Given that the proposed state space is much smaller than previous ones, we compute the pointer's posterior probability exactly instead of approximating it using Monte Carlo PF~\cite{Heydari:2}. In the following, we describe the computational process for the beat state space. The process for the downbeat state space is the same, and is omitted here to save space.

Let $\phi_{k}$ and $y_{k}$ denote the latent state and observation at frame $k$, respectively. Suppose the position posterior $p(\phi_{k}|y_{1:k})$ is already estimated, then a ``predict-update'' iterative procedure can be used to compute the next frame's position posterior $p(\phi_{k+1}|y_{1:k+1})$. First, a one-step-ahead prediction is computed as: 
\begin{equation}
p(\phi_{k+1}|y_{1:k}) = \sum_{\phi_{k}} p(\phi_{k+1}|\phi_{k})p(\phi_{k}|y_{1:k}),
\end{equation}
where $p(\phi_{k+1}|\phi_{k})$ is the state transition probability:
\begin{equation}
    p( \phi_{k+1} |  \phi_{k}) =
      \left\{
        \begin{array}{ll}
            \gamma(\phi_{k}) & \text{if } \phi_{k+1}=1 \\
            1-\gamma(\phi_{k}) & \text{if } \phi_{k+1}=\phi_{k}+1\\ 
            0 &  \text{otherwise } 
        \end{array}
      \right. ,
\end{equation}
where $\gamma(\phi_{k})$ is the normalized jump-back probability for the pointer to jump from position $\phi_{k}$ back to the beat state. The pointer can also move one step to the right with the probability of $1-\gamma(\phi_{k})$, but no other transition is allowed.

Given the tempo range $[T_{min}, T_{max}]$, the jump-back would only happen between states $M(T_{max})$ and $M(T_{min})$, where the former is the least possible number of frames within a beat and the latter is the maximum possible number of frames. 
In other words, $\gamma({\phi_{k}})=0$ for $\phi_{k}\in \{1,2,...,M(T_{max})-1\}$. Also, we set $\gamma(M(T_{min}))=1$ to guarantee that the probability mass never exceeds the last possible state. For the rest of the states i.e., $\phi_{k-1}\in \{M(T_{max}),M(T_{max})+1,...,M(T_{min})-1\}$, jump-back probabilities require updating. 

Then the update step absorbs the newly observed audio frame as follows:
\begin{equation}
p(\phi_{k+1}|y_{1:k+1})= \frac{1}{Z_{k+1}}p(y_{k+1}|\phi_{k+1})p(\phi_{k+1}|y_{1:k}) ,
\end{equation}
where $Z_{k+1}$ is the normalization constant, and the observation likelihood $p(y_{k+1}|\phi_{k+1})$ is defined as: 
\begin{equation}
    p(y_{k+1} | \phi_{k+1}) \propto
      \left\{
        \begin{array}{ll}
            b_{k+1} & \quad \text{if } \phi_{k+1}=1\text{  and  } b_{k+1} \geq T \\
            \epsilon & \quad \text{otherwise}
        \end{array}
      \right. ,
\end{equation}
where the first branch is for the beat state, i.e., the pointer position $\phi_k$ is at the left end of Fig. 1b. We also use a threshold $T$ to omit frames that have too low beat activation $b_{k}$, which is computed from audio features using a neural network~\cite{Heydari:2} or other models. The second branch is for the other states, and a small constant $\epsilon$ is assigned as the likelihood. 



\subsection{Jump-back reward strategy}
\label{ssec:Jump back reward}

In this section, we introduce a method to update the jump-back probability vector. It is noted that jump back technique should not be confused with back tracking operation that is included in some dynamic programming offline inferences such as that of Ellis ~\cite{Ellis:1}. The transition model used in the prediction stage returns a portion of the position probability to the beat state. For the update step, when the beat activation is larger than the threshold, it increases the position probability of the beat state, and decreases that of the other states. 
For $\phi_{k+1}\in \{M(T_{max}):M(T_{min})-1\}$, updating the jump-back probabilities is accomplished through an iterative equation with a forgetting factor $\lambda$ at each frame using an update signal denoted by $\Gamma(\phi_{k+1})$: 
\begin{equation}
\gamma(\phi_{k+1})= \lambda \gamma(\phi_{k})+ (1-\lambda) \Gamma(\phi_{k+1}), \text{where}  
\end{equation}
\begin{equation}
    \Gamma(\phi_{k+1})=
          \left\{
        \begin{array}{ll}
           p(\phi_{k+1}|y_{1:k})-p(\phi_{k+1}|y_{1:k+1})  \ \ \ \text{if }\ b_{k+1} \geq T \\     \\       -(p(\phi_{k}|y_{1:k})-p(\phi_{k+1}|y_{1:k})) \ \ \ \ \ \text{if }\ b_{k+1} < T  \\
            \qquad \qquad \qquad \qquad \qquad \qquad \ \ \ \  \text{and} \ \phi_{k+1}=1 \\
            0 \qquad \qquad \qquad \qquad \qquad \qquad \ \ \ \  \text{otherwise}\\
        \end{array}
      \right. .
\end{equation}
The first branch corresponds to the situation where the $(k+1)$-th frame is likely a beat. It computes the amount of probability changes in the update step, which are probability decreases for non-beat states. This probability change would be informative
in detecting the main loop (tempo), since the higher contrast between before and after update stage indicates that from which state the majority of the probability mass (frame phase) loops back.
The second branch corresponds to the situation where the $k+1$-th frame is not likely a beat. It is the negative amount of probability mass that is jumped back to the beat state in the prediction step. This wrong jump back would have been avoided in an ideal case, and its negative value serves as a punishment for the wrong jumps. The source code and video demos of the implementation are available\footnote{https://github.com/mjhydri/1D-StateSpace}.



\section{Evaluation}
\label{sec:pagestyle}

To demonstrate the capability of the 1D state space, we implemented a causal joint beat and downbeat tracking system using the proposed model. It is important to note that the model does not require tempo and meter prior knowledge and decodes them as well. The beat and downbeat activations are obtained from the pre-trained neural networks in our previous BeatNet work~\cite{Heydari:2}. Also, The states' probabilities and jump-back probabilities are initialized randomly. 
Following the SOFA online methods, we employ the GTZAN dataset to evaluate the performance of the proposed inference model. It is a comprehensive dataset including 1000 excerpts from 10 different music genres. Another reason that makes it more suitable for the evaluation is that it was entirely unseen during the training stage of all reported supervised models.  

Table 1 illustrates the beat and downbeat F-measure performance of several online methods. To demonstrate the efficiency of the proposed model, the processing time of each model is reported as well. The reported numbers are the average computational time for 30-second music excerpts of the GTZAN dataset. Note that we measure the speed of all methods on the same Windows machine with an AMD Ryzen 9 3900X CPU and 3.80 GHz clock. For DLB~\cite{Heydari} and BeatNet~\cite{Heydari:2}, their paper-reported default number of particles is used, which is 1000 and 1750, respectively. 

Among all of the methods in Table 1, the new model and BeatNet~\cite{Heydari:2} are joint beat and downbeat tracking approaches, and the rest are only beat tracking models. Also, except IBT~\cite{Oliveria} and Aubio~\cite{Brossier}, all the other models are supervised, and they leverage deep neural networks to extract beat and/or downbeat activations.     
Table 1 demonstrates that the proposed model can deliver the same beat tracking F-measure performance as the BeatNet~\cite{Heydari:2}, which is the SOFA online approach. However, its downbeat performance is moderately lower than that of the BeatNet. The primary point is that using the proposed 1D state space and the jump-back reward technique leads to a more than 30x speedup than BeatNet, making it much more suitable for large scale industrial usages.

Another interesting observation is that Aubio~\cite{Brossier} is even faster than the proposed model. The reason is that it is a classical signal processing model and it does not use Bayesian temporal decoding at the inference stage. Its main drawback, however, is its performance which is the lowest and suffering from the common issues of the sliding window approaches~\cite{Oliveria}. DLB~\cite{Heydari}, on the other hand, addresses the sliding window issues and delivers much better results, but it is the slowest model.  
Finally, comparing the proposed model with Böck FF~\cite{Bock:3,Bock:4}, which also performs exact inference instead of using a sampling approach,
we see that the proposed method achieves a similar F-measure on beat tracking and a 7.5x speedup, even though the proposed model is a multi-task approach that also performs downbeat tracking. 

\begin{table}[t]
  \begin{center}
      \caption{Performance and speed comparison of several online beat and downbeat tracking models on the GTZAN.\\}
    \begin{tabular}{lccc}
        \hline
        \small\textit{Method} & \small\textit{F-Measure} & \small\textit{F-Measure} & \small\textit{Comp. Time} \\
         & \small\textit{Beats} & \small\textit{Downbeats}   &\small\textit{(Seconds)} \\
        \hline
        \small Aubio \cite{Brossier} & \small 57.09 & \small --- & \small 0.1 \\
        \small BeatNet \cite{Heydari:2} & \small \small{75.44} & \small \underline{46.49} & \small 8.87\\
        \small Böck ACF \cite{Bock:2} & \small 64.63 & \small --- & \small 7.01\\
        \small Böck FF \cite{Bock:3} & \small 74.18 & \small --- & \small 2.19\\
        \small DLB \cite{Heydari} & \small 73.77 & \small --- & \small 21.30\\
        \small IBT \cite{Oliveria} & \small 68.99 & \small --- & \small 4.89\\
        \small \textbf{1D state space}  & \underline{76.48} & \small 42.57 & \small 0.29\\
        \hline
    \end{tabular}
\vspace{-0.2 in}
    \label{tab:results}
  \end{center}
\end{table}

\section{Conclusion and Future works}
\label{sec:conclusion}
In this paper, we introduced a novel 
semi-Markov
state space and a jump-back reward technique to reduce the computational cost of music rhythmic analysis tasks. This new state space is much smaller than the previous efficient space in the literature. We also implemented an online model to infer several music rhythmic parameters jointly. We showed that using the new state space along with the new inference process delivers the SOFA results for beat tracking and comparable results for downbeat tracking with a drastically faster speed.




\bibliographystyle{IEEEbibvv}
\bibliography{refs}

\end{document}